\documentstyle[epsf]{mn}

\title[Optical/Infrared observations of RX J1914+24]{Optical/Infrared
spectroscopy and photometry of the short period binary RX J1914+24
\thanks{Based (in part) on data collected at Subaru Telescope, which
is operated by the National Astronomical Observatory of Japan.}}

\author[G. Ramsay et al.]{Gavin Ramsay$^{1}$, Kinwah Wu$^{2,1}$,
      Mark Cropper$^{1}$, Gary Schmidt$^{3}$, Kazuhiro Sekiguchi$^{4}$ \and
      Fumihide Iwamuro$^{5}$, Toshinori Maihara$^{5}$\\
$^{1}$Mullard Space Science Laboratory, University College London, 
      Holmbury St. Mary, Dorking, Surrey RH5~6NT\\
$^{2}$Research Centre for Theoretical Astrophysics, School of Physics,
      University of Sydney, NSW 2006, Australia\\
$^{3}$Steward Observatory, The University of Arizona, AZ 85721, USA\\
$^{4}$National Astronomical Observatory of Japan, 
      650 North A'ohoku Place, Hilo, Hawaii 96720, USA\\
$^{5}$Department of Physics, Kyoto University, Kitashirakawa, 
      Kyoto 606-8502, Japan}

\date{}

\begin{document}
\def\Mdot{\hbox{$\dot M$}}
\def\Msun{\hbox{$M_\odot$}}
\def\Rsun{\hbox{$R_\odot$}}
\outer\def\gtae {$\buildrel {\lower3pt\hbox{$>$}} \over 
{\lower2pt\hbox{$\sim$}} $}
\outer\def\ltae {$\buildrel {\lower3pt\hbox{$<$}} \over
{\lower2pt\hbox{$\sim$}} $}
\def\rchi{{${\chi}_{\nu}^{2}$}} 
\newcommand{\ergss} {ergs s$^{-1}$} 
\newcommand{\pcmsq} {cm$^{-2}$}

\maketitle

\begin{abstract}

We present observations of the proposed double degenerate polar
RX~J1914+24. Our optical and infrared spectra show {\it no} emission
lines. This, coupled with the lack of significant levels of
polarisation provide difficulties for a double degenerate polar
interpretation. Although we still regard the double degenerate polar
model as feasible, we have explored alternative scenarios for
RX~J1914+24. These include a double degenerate algol system, a neutron
star-white dwarf pair and an electrically powered system. The latter
model is particularly attractive since it naturally accounts for the
lack of both emission lines and detectable polarisation in RX
J1914+24. The observed X-ray luminosity is consistent with the
predicted power output. If true, then RX J1914+24 would be the first
known stellar binary system radiating largely by electrical energy.

\end{abstract}


\section{Introduction}

RX J1914+24 was discovered during the course of the {\sl ROSAT}
all-sky survey and was found to show a modulation in X-rays on a
period of 9.5 min (Motch et al 1996).  In further observations,
Cropper et al (1998) detected only the 9.5 min period which they
suggested was the binary orbital period.  This implied a very small
binary dimension. This together with the X-ray light curve, which was
consistent with zero for half the 9.5 min period, led them to propose
that the system is a polar (those magnetic cataclysmic variables --
mCVs -- in which the accreting white dwarf has a magnetic field strong
enough to synchronise its spin period with the binary orbital period).
This would imply that the 9.5 min period seen in X-rays was both the
spin period of the accreting white dwarf and also the binary orbital
period.  For an orbital period this short the secondary star could not
be a main sequence star as in the case of ordinary mCVs.  It could,
however, be a He white dwarf.  If confirmed, RX~J1914+24 would be the
first binary system in which both components are white dwarfs and
magnetically synchronised -- a double degenerate polar.  Furthermore,
its orbital period would be the shortest of any known stellar binary
system.

Observations by Ramsay et al.\ (2000) gave supporting evidence for the
double degenerate polar interpretation.  These include: discovery of
the optical counterpart (in the $I$ band) in which only one period was
seen (the 9.5 min period seen in X-rays); no periods other than the
9.5 min period in further X-ray data; an X-ray spectrum which is very
soft and typical of polars; large variations in its long-term X-ray
light curve which are typical of polars, and infrared colours which
are not consistent with that of a main-sequence star.  Ramsay et al.\
(2000) also found that the peak intensity in the X-ray and $I$ band
light curves were approximately anti-phased. They interpreted the
X-ray flux as originating from the accretion region on the white
dwarf, while the $I$-band flux originated from the irradiated face of
the mass donating white dwarf.  This was consistent with the
observation that the $I$-band flux was not polarised.

In this paper we present further observations of RX~J1914+24. We
discuss these results in relation to the double degenerate polar
model of Cropper et al (1998) and Ramsay et al (2000). We also
consider several alternative models for RX~J1914+24.

\section{Optical/Infrared Imaging}

\subsection{$J$-band imaging using Subaru}

\begin{figure}
\begin{center}
\setlength{\unitlength}{1cm}
\begin{picture}(8,7.5)
\put(1.,-18){\includegraphics{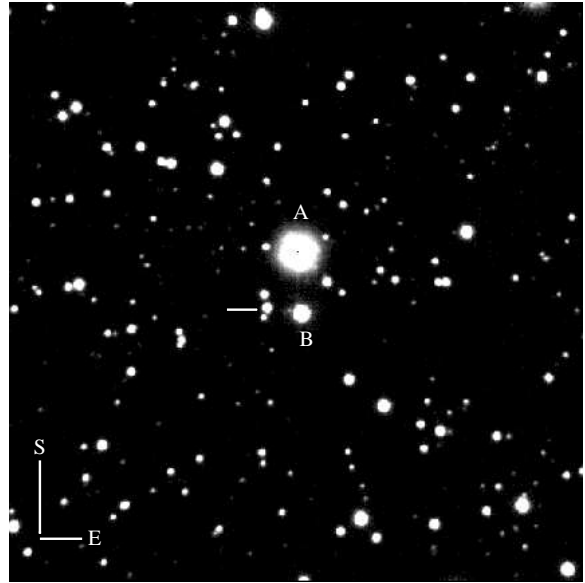}}
\end{picture}
\end{center}
\caption{ A 60-min $J$-band image of the field of RX~J1914+24 taken
using the 8-m Subaru telescope on Hawaii.  Stars A \& B are shown in
the images of Motch et al.\ (1996) and Ramsay et al.\ (2000) and are
separated by 7~arcsec.  Star A has $V\sim$15.5 (from data obtained at
the NOT) and $J\sim$12.1 (from UKIRT data, Ramsay et al.\ 2000) and
was saturated in each 5-sec exposure.  The optical counterpart of
RX~J1914+24 is shown by the white marker.  }
\label{finding} 
\end{figure}

RX~J1914+24 was observed on 1999 June 6 in the $J$ band using the 8-m
Subaru telescope on Hawaii and the Cooled Infrared Spectrograph and
Camera (CISCO) in its imaging mode.  Each exposure was 5~sec in
duration, and after every 12 exposures an offset was applied to the
telescope so that a flat field could be obtained.  The observation
lasted $\sim$1 hour.  Conditions were good and the seeing was
$\sim$0.4 arc sec.  An image using all the frames (each frame was
shifted in position so the stellar images were coincident) is shown in
Figure \ref{finding}.

To make an accurate determination of the position of the optical
counterpart of RX~J1914+24, we matched the positions of stars in the
Subaru image with those in the Digitised Sky Survey.  The accurate
positions of these stars were then taken from the USNO A1.0 catalogue.
We then determined the position of RX~J1914+24 using the astrometry
package {\tt ASTROM} (Wallace 1998): $\alpha_{2000}$ =
19$^{h}$~14$^{m}$~26.1$^{s}$, $\delta_{2000}$ =
+24$^{\circ}$~56${'}$~43.6$^{''}$.  The uncertainty on this position
is 0.4$^{''}$.  It is consistent with the X-ray position reported by
Cropper et al.\ (1998).

Aperture photometry was performed on the images using {\tt PHOTOM}
(Eaton, Draper \& Allan 1999).  Differential photometry was obtained
using various comparison stars in the field.  Figure \ref{fold} shows
the differential light curve of the optical counterpart to RX~J1914+24
binned and folded on the ephemeris of Ramsay et al.\ (2000).  (The
ephemeris of Ramsay et al.\ (2000) is sufficiently precise to phase
the Subaru data with the original X-ray and $I$ data to within 0.02
cycles).  The shape and phasing of the light curve is similar to that
of the $I$-band light curve shown in Ramsay et al.\ (2000) (and also
shown here for comparison) although the rise to maximum is less
rapid. The amplitude is greater in $J$ (0.16 mag peak-to-peak)
compared to $I$ (0.07 mag). By comparison, the $J$-band data obtained
using UKIRT (Ramsay et al.\ 2000) was of too low a signal to noise for
a significant modulation to be apparent.

\subsection{$BVRI$ imaging using NOT}
\label{bvri}

Observations were obtained on the nights 1999 July 6--8 using the
2.5-m Nordic Optical Telescope (NOT) on La Palma, and the Andalucia
Faint Object Spectrograph (ALFOSC) was used in its imaging mode.
The conditions were photometric on the first two nights.  The seeing
was typically 1$^{''}$.  To obtain circular polarimetry data a
1/4-waveplate was inserted into the optical path: this split the
light into both polarised beams on the Loral 2k$\times$2k CCD so
that 2 images of each star were recorded.  The instrument was
orientated in such a way that images of field stars did not overlap
with any of the stars of interest.  The images were bias-subtracted
and flat-fielded in the usual way.  On the first night images were
mainly taken in the $R$ band (20-sec exposures) and the second night
the $V$ band (40-sec exposures).  The CCD was windowed to reduce the
readout time.

In addition to the extended sequence of $V$- and $R$- band images, we
also made a sequence of $B, V, R$ and $I$ images and a standard star
at the beginning of each night.  This allowed us to deduce magnitudes
for RX~J1914+24 across the broad optical/infrared band -- see Table
\ref{allphot}.  In the $I$ band RX~J1914+24 was found to be
$\sim$4.8~mag fainter than star `A' -- the same as it was when
observed in June 1998 (Ramsay et al.\ 2000).

\begin{figure}
\begin{center}
\setlength{\unitlength}{1cm}
\begin{picture}(8,12.5)
\put(-0.5,-1.){\includegraphics{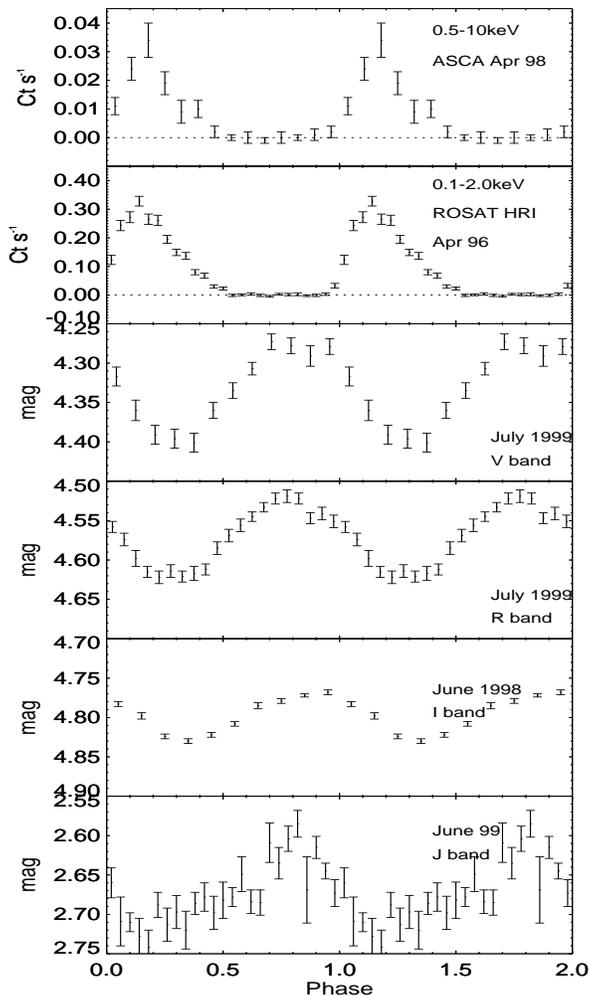}}
\end{picture}
\end{center}
\caption{ The light curve of RX~J1914+24 as a function of energy
folded on the 9.5 min period.}
\label{fold} 
\end{figure}

Profile fitting photometry was carried out on stars in the field using
{\tt DAOPHOT} (Stetson 1992). Since the profiles of the two polarised
beams were different, two point spread functions had to be constructed
We show the $V$ and $R$ differential light curves (using one of the
polarised beams) of the optical counterpart of RX~J1914+24 folded and
binned on the ephemeris of Ramsay et al.\ (2000) in Figure \ref{fold}.
Both light curves resemble those in the $I$ and $J$ bands with
their peak intensities occurring $\sim$0.3--0.4 cycles before the
maximum in X-rays.  The peak to peak amplitude is greatest in $V$:
0.13 mag in $V$, 0.10 in $R$ and 0.07 in $I$.

We note the presence of a dip at $\phi\sim$0.85 which is seen in the
$V$, $R$ and $J$ bands.  (The time resolution of the $I$ band data is
poorer than the other bands, which probably prevents us detecting it
in this band).  We would not regard this dip as significant if
observed in only one band, but its presence in 3 bands (taken at
different epochs) suggests that it is indeed significant.

The presence of a 1/4 waveplate and calcite block in the optical path
allowed us to search for circular polarisation in RX~J1914+24.  The
instrumental polarisation was found to depend on the position of the
stellar image on the CCD.  To make a first order correction for
instrumental polarisation we made the assumption that star `A' has no
intrinsic circular polarisation (cf.\ Figure \ref{finding}).  We
applied this correction to the polarised light curves of star `B' and
RX J1914+24 and then folded them on the ephemeris of Ramsay et al.\
(2000).  No significant polarisation was found in RX~J1914+24 in
either the $V$ or $R$ bands.  To test that the instrument could detect
polarisation, observations of the polar QQ Vul were made at the end of
the first and second nights.  It was found that significant circular
polarisation was detected, with the same polarity and similar degree
of polarisation detected in the observations of Cropper (1998).

\begin{table}
\begin{tabular}{llll}
\hline
Band&Mean     &Dereddened&Dereddened Flux\\
    &Magnitude&Magnitude&erg cm$^{-2}$ s$^{-1}$ \AA$^{-1}$\\
\hline
B&21.1$\pm$0.15&16.9&1.1$\pm0.2\times10^{-15}$\\
V&19.9$\pm$0.10&16.9&6.3$\pm0.9\times10^{-16}$\\
R&19.2$\pm$0.10&17.0&3.6$\pm0.3\times10^{-16}$\\
I&18.6$\pm$0.10&17.2&1.6$\pm0.1\times10^{-16}$\\
J&17.3$\pm$0.10&16.5&8.3$\pm0.8\times10^{-17}$\\
H&17.4$\pm$0.15&16.9&2.0$\pm0.3\times10^{-17}$\\
K&17.1$\pm$0.15&16.8&7.8$\pm1.2\times10^{-18}$\\
\hline
\end{tabular}
\caption{The apparent and dereddened magnitudes and fluxes of 
RX~J1914+24 assuming an extinction of $A_{V}$ = 3.0 mag.  
The $B, V, R,
I$ estimates were made in July 1999 while the $J, H, K$ magnitudes
were made in July 1998 (when the $I$ magnitude was within 0.1 mag of
that in July 1999). The error on the mean magnitudes includes the
error in placing them onto the standard system. The error on the
dereddened flux does not include the uncertainty on the interstellar 
extinction.}
\label{allphot}
\end{table}  

\section{Spectroscopy} 

\subsection{The optical spectrum}
\label{optspec}

Central to their double degenerate polar interpretation, Cropper et
al.\ (1998) noted that the shortest orbital period a binary system
could have with a main sequence secondary is $\sim$80~min (see Ritter
1986).  However, with a He degenerate secondary, the binary orbital
period could lie in the range $\sim$6--50~min.  A crucial test is the
characteristics of the optical/infrared spectrum of RX~J1914+24 -- if
hydrogen lines were to be detected, then this would cast considerable
doubt upon the double degenerate polar interpretation.

\begin{figure}
\begin{center}
\leavevmode
\setlength{\unitlength}{1cm}
\begin{picture}(8,7.5)
\put(-2,-2){\includegraphics{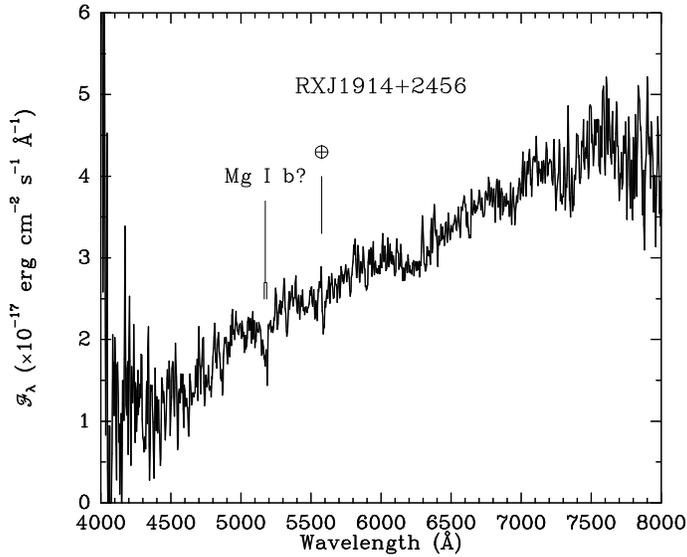}}
\end{picture}
\end{center}
\caption{ The optical spectrum of RX~J1914+24 taken with the 2.3-m
telescope at Steward Observatory at Kitt Peak, no correction for
reddening has been applied. No significant emission
lines are detected.}
\label{gary} 
\end{figure}

Spectra were taken with the 2.3-m telescope at Steward Observatory at
Kitt Peak on the nights of 1998 Sep 24 and 1998 Nov 15, under
conditions of moderate ($\sim1.0-1.5^{"}$) seeing with a slit of 3"
and 2" width, respectively.  The object could be discerned from
surrounding nearby stars on the guider TV.  The total exposure time
was 5400~sec, using the CCD spectropolarimeter developed by Schmidt et
al.\ (1992).  Resolution was about 12 \AA.  The resulting spectrum
(Figure \ref{gary}) is very red and, remarkably, {\it no} emission
lines are detected.  The only possible line is an absorption line at
$\sim$5200\AA \ which could be due to Mg I (5173/5184 \AA).  This
feature was only observed on one of the two nights so this feature
must be regarded as tentative.

For the data taken in Sept 1998, the instrument was configured for
circular spectropolarimetry. Using the whole spectral range we find
a formal result of V/I = 0.14~percent +/- 0.29~percent (1 sigma), for a
3~sigma upper limit of 0.9~percent.  This is consistent with our
imaging polarimetry data.  There are no significant polarization
features in the spectrum from 4000--8000 \AA.

\subsection{The $K$-band Infrared Spectrum}
\label{irspec}

\begin{figure}
\begin{center}
\setlength{\unitlength}{1cm}
\begin{picture}(8,5)
\put(-1.5,-28.5){\includegraphics{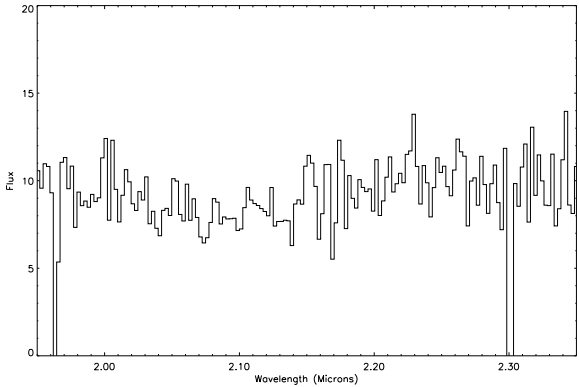}}
\end{picture}
\end{center}
\caption{ The $K$-band infrared spectrum of RX J1914+24 taken with the
UKIRT telescope.  The flux scale is arbitrary.}
\label{ukirtspec} 
\end{figure}

We obtained a $K$-band infrared spectrum of RX~J1914+24 on 1999 August
14 using UKIRT on Hawaii as part of the service program.  The detector
was CGS4.  A 30-sec exposure was taken at each detector position, then
the detector was moved by half a pixel over a distance of 2~pixels.
Each frame was therefore a total of 120~sec.  The combined exposure
was 104~min in total.  The bright star BS7280 was observed to correct
for telluric features.  The images were reduced in the standard manner
using {\tt FIGARO} procedures.  Because of variations in the seeing
and water content in the atmosphere during the observations, it is not
possible to obtain reliable flux calibrated spectra, and therefore the
flux scale is arbitrary.  The resulting spectrum is shown in Figure
\ref{ukirtspec}.  Although the signal to noise is low, no obvious
emission lines of either H or He are present.

In the 1.9--2.4-$\mu$m wavelength range, several emission lines have
been observed in the spectra of mCVs with red-dwarf secondaries
(Dhillon \& Marsh 1995; Dhillon et al.\ 1997; Dhillon 1998; Dhillon et
al.\ 2000).  These include the prominent emission lines He I
(2.059$\mu$m) and Bracket $\gamma$ (2.166$\mu$m).  Dhillon et al.\
(2000) also show the spectrum of the non-magnetic, double degenerate
system, GP Com, which exhibits the prominent He I line at
2.059~$\mu$m.  These spectra were also obtained using UKIRT and CGS4.
Although there is no known $K$-band magnitude for GP Com in the
literature, its $V$-band magnitude is $\sim$16, compared to $\sim$20
when we observed RX~J1914+24 in July 1999 (the month previous to our
UKIRT spectrum).  The exposure of the GP Com spectra (24 min) is a
factor of 4.3 shorter than that of our RX~J1914+24 spectrum, which is
equivalent to reaching 1.6 mag fainter.  If the $V$-band magnitude
difference between GP Com and RX~J1914+24 is roughly comparable in the
$K$ band, this would suggest that the lines may be hidden in the
noise.

\section{The extinction towards RX J1914+24}
\label{extinc}

Haberl \& Motch (1995) fitted the {\sl ROSAT} X-ray spectrum of RX
J1914+24 with an absorbed blackbody spectrum and obtained a column
density of 1.0$\pm0.1\times10^{22}$ cm$^{-2}$. From this they derived
an extinction of $A_{V}$=5.6 using the $N_{H}, E_{B-V}$ relationship
(eg Predehl \& Schmitt 1995) and $A_{V}=3E_{B-V}$. Using this value
for the extinction, Ramsay et al (2000) found that the extinction
corrected $I, J, K,$ \& $H$ fluxes could not be well fitted with a
blackbody. We now have observations extending down to the $B$ band
which allows us to re-examine the extinction towards RX J1914+24.

The $BVRI$ band data shown in Table 1 were taken on 6--8 July 1999
while the $JHK$ were taken in 2--3 July 1998. In addition $I$ band
data were taken on 25--27 June 1998 which show that RX J1914+24 was
approximately the same magnitude in June 1998 as July 1999 (to within
$\sim$0.1 mag). This suggests that we can combine the $BVRIJHK$ data
with a reasonable degree of confidence.

\begin{figure}
\begin{center}
\setlength{\unitlength}{1cm}
\begin{picture}(8,12)
\put(-1.5,4.6){\includegraphics{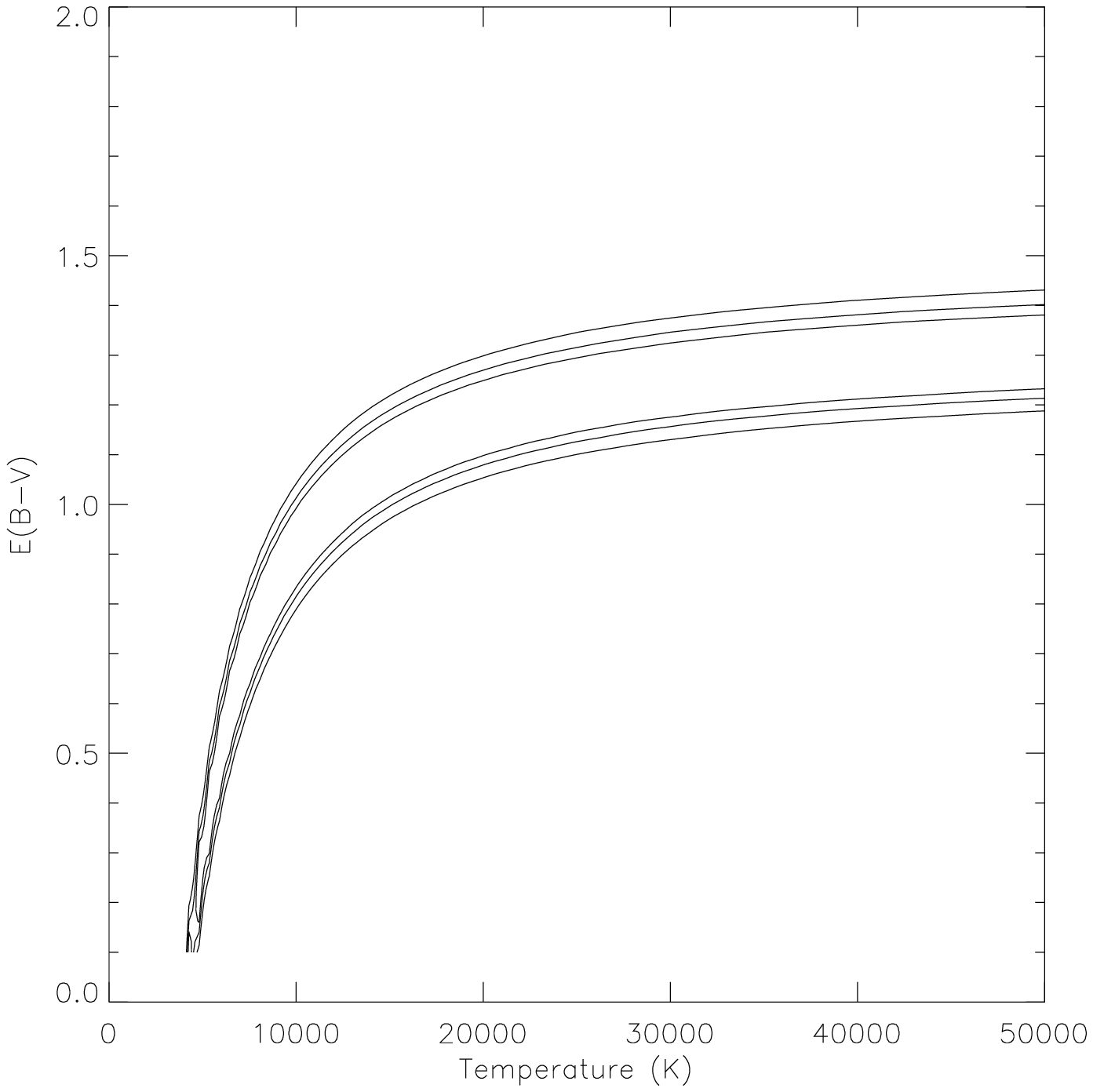}}
\put(-1.5,-1.9){\includegraphics{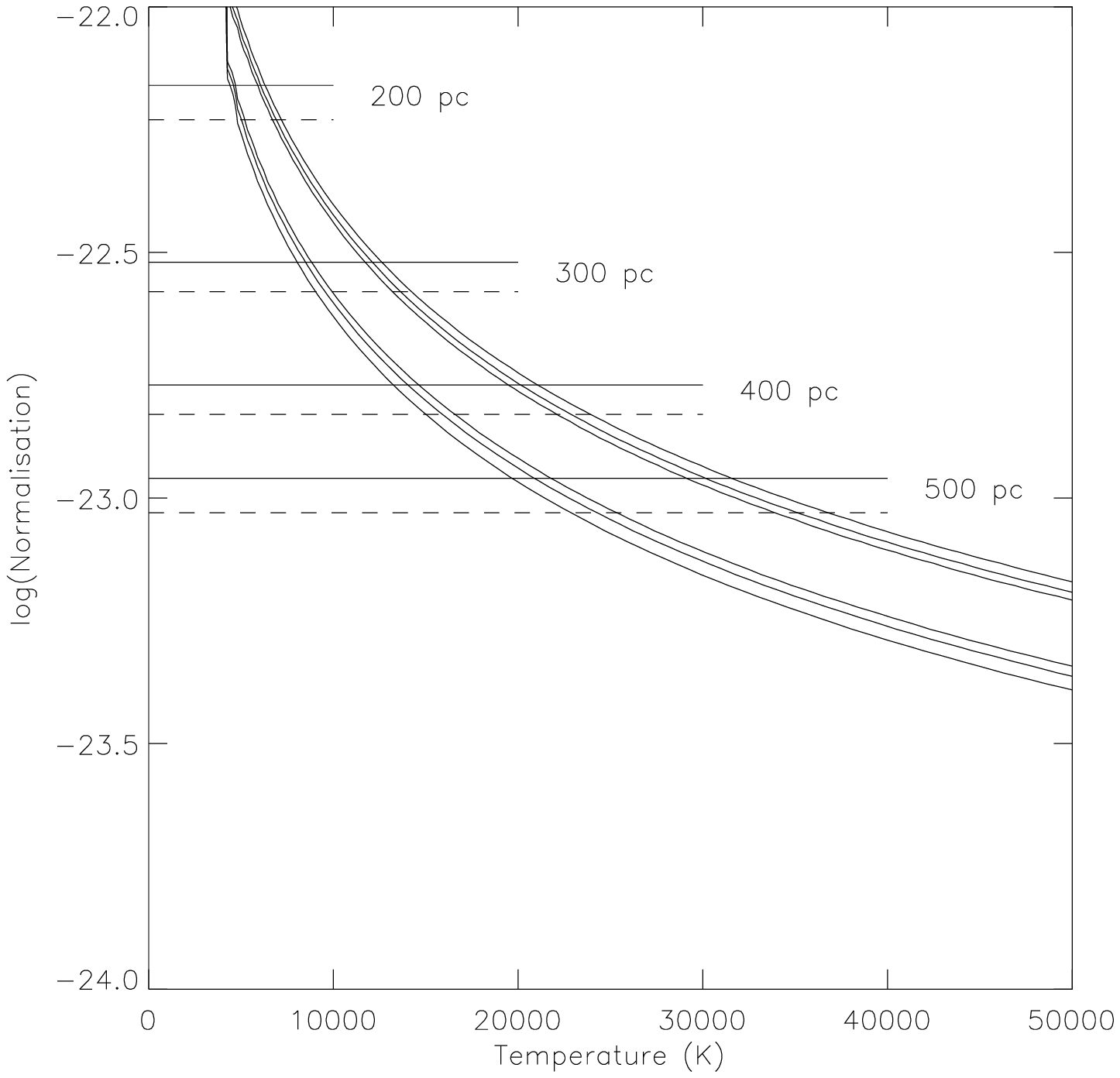}}
\end{picture}
\end{center}
\caption{The temperature, extinction and normalisation (=Area white
dwarf/2distance$^{2}$) derived using a blackbody fit to the $BVRIJHK$
data. The 68, 90 and 99 per cent confidence contours are shown.  The
solid lines are for a white dwarf of radius appropriate for a Roche
lobe filling secondary (a 0.08\Msun white dwarf) and the dashed lines
are for a slightly underfilling its Roche lobe (a 0.1\Msun white dwarf).}
\label{dered} 
\end{figure}

We fitted a blackbody to the optical-infrared data assuming an area of
a Roche lobe filling secondary (a 0.08\Msun white dwarf) and one which
just underfills its Roche lobe (a 0.1\Msun white dwarf). We caution
that for such low masses, the reliability of the Nauenberg (1972)
mass-radius relationship for white dwarfs is not known. The free
parameters were the temperature of the blackbody; the extinction and
the normalisation (or distance). For the extinction we take the
interstellar reddening law given in Zombeck (1990). We show in Figure
\ref{dered} the confidence contours in the Temperature--$E_{B-V}$ and
Temperature--normalisation planes.

We find that a wide range of temperatures are consistent with the
data. Single DA white dwarfs have a mean temperature of $\sim$10000K,
while accreting white dwarfs usually have photospheric temperatures in
the range $\sim$10000--30000K (Sion 1998). For a temperature of
10000K, $E_{B-V}\sim0.85, A_{V}=2.6$, while for a temperature of
30000K, $E_{B-V}\sim1.25, A_{V}=3.9$. For a temperature in the range
10000--30000K the distance is $\sim$200--500pc. To match the
extinction derived by Haberl \& Motch (1995), the temperature would
have to be much greater than 50000K. From similar arguments, Ramsay et
al (2000) estimate a distance of 100--400pc.

For an extinction of $A_{V}$=3.0, the implied unabsorbed luminosity
over the 4000--25000 \AA\hspace{1mm} range is $\sim$ $2\times10^{31}
d^{2}_{100}$ \ergss where $d^{2}_{100}$ is the distance in units of
100pc.

Since we derive a lower absorption than Haberl \& Motch (1995) we
extracted the {\sl ROSAT} PSPC spectrum from the data archive. Rather
than binning the spectrum into 6 bins, we binned it into 52 bins. We
fitted the spectrum with an absorbed blackbody and find that the
temperature and absorption column is closely correlated and the
absorption is not closely constrained. We find that if we fix the
extinction at $A_{V}$=3.0 (=5.4$\times10^{21}$ cm$^{-2}$) we obtain a
good fit: \rchi=1.05 (50 dof), $kT=$56eV. This is consistent with the
fits to the {\sl ASCA} SIS spectrum (Ramsay et al 2000). The
phased-averaged bolometric X-ray luminosity is
$1.2^{+4.1}_{-0.3}\times10^{33} d^{2}_{100}$ \ergss.

\section{Amplitude variations in Optical/Infrared}
\label{ampvar}

In Figure \ref{fold} we show the $V$, $R$, $I$, $J$ photometry folded
and binned on the 9.5 min orbital period.  The peak to peak amplitude
variation is 0.13, 0.10, 0.07 and 0.16 mag respectively. In the double
degenerate model of Ramsay et al (2000) the optical/infrared
modulation is due to the secondary star which is irradiated by the
accretion sites on the primary. We can make a very crude estimate of
the temperature difference between the irradiated and non-irradiated
hemispheres of the secondary star, if we assume they have equal areas
and the emission can be approximated by a blackbody.  In reality
neither assumption is likely to be an accurate approximation, as the
emission is probably characterised by a blackbody component together
with another component which is due to irradiation. Other effects,
such as limb darkening will also complicate matters.  Assuming a mean
temperature for the secondary of T=20000K we find that the observed
amplitude variations require a temperature difference of
$\sim$1000--2000K between the irradiated and non-irradiated parts of
the secondary.

\section{The nature of the short dip seen in the optical}
\label{dip}

In \S \ref{bvri} we concluded that the short dip seen at phase
$\phi\sim$0.85 was significant.  Using the $J$-band data, which has
the highest time resolution, we find that the width of the dip is
$\Delta\phi\sim$0.03 cycles.  The phasing of the optical/infrared
light curve with respect to the X-ray curve suggests that it could be
due to a stream which partially obscures the irradiated surface of the
secondary. Since the dip is so narrow, the stream should have a small
diameter.

Suppose RX J1914+24 consists of a magnetic white dwarf accreting material
from its companion. For a dipole magnetic field, one can define a
magnetospheric radius, $r_{\mu}$, at which point the magnetic pressure
balances the ram pressure of the accretion flow. The magnetospheric
radius takes the form:

\begin{equation} 
   r_{\mu}\ =\  5.1\times10^{8} 
   M_{1}^{-1/7} \Mdot_{16}^{-2/7} \mu_{30}^{4/7}~{\rm cm} 
\end{equation} 
which gives $r_{\mu}=1.2\times10^{10}$ cm for $M_{1}$=1.0\Msun (giving
$R_{1}=5.5\times10^{8}$ cm), $B=$5 MG and \Mdot=1$\times10^{16}$ g
s$^{-1}$.  This is comparable to the binary separation of
$a\sim1.1\times10^{10}$~cm for the same orbital parameters.  This
suggests that the stream gets coupled by the magnetic field very close
to the $L_{1}$ point.

The azimuthal angular span of the accretion stream is approximately
constant from the $L_{1}$ point towards the primary and hence the
width of the dip seen in the optical can be estimated if we can
determine the radius of the stream at the $L_{1}$ point.
The radius of the stream at the $L_{1}$ point, $r_{_{L1}}$, is
   roughly given by
\begin{equation} 
   r_{_{\rm L1}} \sim \sqrt{H R_{2}} \,
\end{equation} 
   where $H$ is the scale height of the unperturbed atmosphere, 
   given by 
\begin{equation} 
   H \sim \frac{kT_{2}R_{2}^{2}}{\mu_{m}m{_H}GM_{2}} \ ,
\end{equation} 
and $T_{2}$ is the surface temperature of the secondary star,
$\mu_{m}$ is the mean molecular mass, $M_{2}$ is the mass of the
secondary (Pringle 1985). In our double degenerate polar
interpretation, we have $M_{2}$=0.08\Msun, giving
$R_{2}=2\times10^{9}$~cm from the Nauenberg (1972) mass-radius
relationship for white dwarfs. For a helium atmosphere, $\mu_{m}$=4
and for $T_{2}=10^{4}$ K, we obtain $r_{\rm L1}=1.1\times10^{7}$ cm.
This would result in a dip lasting $\sim2\pi/1000\sim$0.006 cycles
which is of the right order of magnitude for the observed duration of
the dip.

\section{Unsolved Mysteries} 

Cropper et al.\ (1998) suggested that RX~J1914+24 is a double
degenerate polar to explain the lack of a second period in the {\sl
ROSAT} data and the shape of the X-ray light curve. The more extensive
study of Ramsay et al.\ (2000), added weight to this argument: the
X-ray spectrum is typical of polars; a main sequence secondary star
was found to be inconsistent with the infrared colours; and a more
detailed search of the X-ray and optical data continued to reveal only
a single period.

At this point it maybe useful to briefly summarise the arguments that
support the presence of mass transfer in RX~J1914+24.  The fact that
we observe a modulation in X-rays and the optical is not by itself
indicative of mass transfer: for instance, a modulation is seen in the
non-interacting white dwarf RE~J0317$-$853 due to its rotation
(Barstow et al.\ 1995).  On the other hand, since the optical and
X-ray light curves are out of phase (Figure \ref{fold}) it is much
more likely that it is indeed an interacting binary.  Further, Ramsay
et al.\ (2000) found a large variation in the long-term X-ray light
curve which can be explained in terms of mass-transfer variations.
These features support the double degenerate polar model. However, in
this model we would also expect to detect optical polarisation and
emission lines -- which we do not detect. We now go on to address these
issues.

\subsection{The problem of no polarisation} 
\label{epoln}

$I$-band polarimetry by Ramsay et al.\ (2000) failed to detect
significant circular polarisation in the system. The anti-phasing of
the $I$-band and X-ray light curves indicates that the optical flux is
unlikely to originate from cyclotron emission. Ramsay et al.\ (2000)
concluded that the magnetic field of the primary must be sufficiently
high for the cyclotron emission to peak in the UV, or sufficiently low
for it to be at infrared wavelengths. Alternatively, they suggested
that no shock formed owing to the small dimensions of the system. Our
$V$ band polarimetric and spectropolarimetric observations presented
in this paper extend the non-detection of circular polarisation to
shorter wavelengths, so that the high-field option above can be
considered less tenable. Unfortunately no infrared circular
polarimetry is available to test the low field option.

\subsection{The problem of no emission lines}
\label{elines}

A second difficulty with the double degenerate polar model is that we
find that {\it no} detectable emission lines.  A general (indeed
almost defining) feature of accreting binary stars is the presence of
emission lines.  In the case of AM CVn systems (non-magnetic
interacting double-degenerate binary systems) the donor is a low-mass
helium white dwarf.  In these systems, no hydrogen lines are visible,
but strong helium lines are present (eg Marsh, Horne \& Rosen 1991).

If mass transfer does occur, the lack of emission lines may be due to
continuum emission from the heated face of the secondary which far
exceeds the flux from any line emission from the accretion stream.
Assuming the temperatures of the heated secondary and the accretion
stream are approximately similar, then the flux ratio from the
secondary and the stream is simply proportional to the relative
emitting areas.  For the secondary, the area is $\pi R_{2}^{2} \sim 
10^{19}$ cm$^{2}$.  Crudely, the projected area of the stream is
$\sim r a$ where $r$ is the radius of the stream and $a$ the orbital
separation.  In \S \ref{dip} we estimate the radius of the stream as
$\sim 10^{7}$ and $a=10^{10}$ cm (Ramsay et al 2000),
giving an area of $10^{17}$ cm$^{2}$.  The flux ratio is therefore
$\sim$100.  For a hot accretion stream, say $T>$30000 K, line emission
would be suppressed and therefore the flux ratio of 100 must be
regarded as a lower limit.  Although the double degenerate polar model
predicts weak optical emission lines to be present, we will require a
spectrum of much higher signal to noise than the spectra obtained in
\S 2 to test this.

\section{Alternative models} 

Despite the success of the double-degenerate polar model in explaining
many of the characteristics of the system, the absence of both
detectable circular polarisation and line emission requires some
explanation. The arguments discussed earlier indicate that it is
possible to explain the absence of these features, but only by
adopting particular values for the magnetic field and that the
emission lines are so weak that we did not detect them in our
relatively low signal to noise spectra. We have therefore explored
whether there may be alternative models for RX J1914+24 which would
explain its characteristics more naturally. We consider below three
alternatives: a double-degenerate algol model, a neutron
star-white dwarf pair and an electrically driven model.

\subsection{Double-degenerate Algol Model}

In normal Algol systems, the accretor is an early main sequence star,
the donor either a late type giant or sub-giant star. For relatively
short period systems (less than $\sim$6 days), the radius of the
primary, $R_{1}$, is large enough compared to the minimum distance
between the stream and the center of the primary ($R_{\rm min}$), that
the accretion stream will impact directly onto the primary without
forming an accretion disc.

\begin{figure}
\begin{center}
\setlength{\unitlength}{1cm}
\begin{picture}(8,8)
\put(-11.5,-1.){\includegraphics{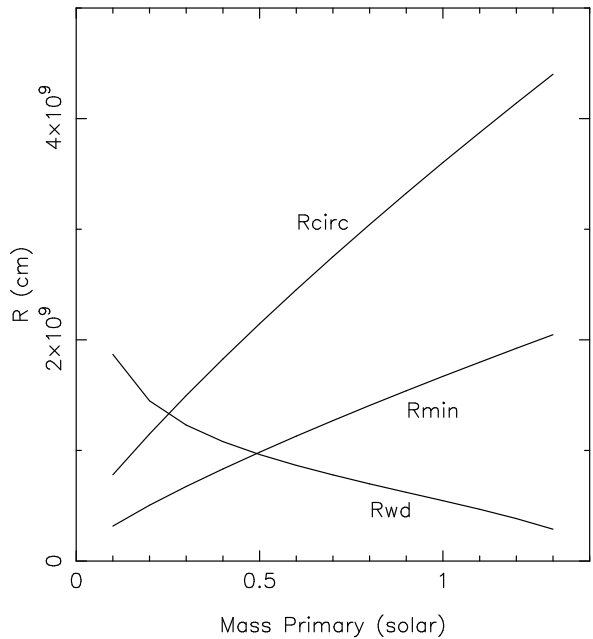}}
\end{picture}
\end{center}
\caption{For a range of masses for the white dwarf primary we show the
radius of the primary ($R_{wd}$), the circularisation radius ($R_{circ}$)
and the distance of minimum approach of the stream to the primary 
($R_{min}$). For $R_{wd}>R_{min}$ the stream will impact directly onto
the primary as in the case for short period Algol systems.}
\label{algol} 
\end{figure}

Here we consider the case when both the primary and secondary are
white dwarfs with negligible magnetic fields. We show in Figure
\ref{algol}, for a range of combinations of $M_{1}$, $R_{\rm min}$ and
$R_{1}$.  We also show for interest, the radius from the primary at
which a disc starts to form -- the circularisation radius $R_{\rm circ}$.
Figure \ref{algol} shows that for $M_{1}<$0.5\Msun the stream will
impact directly onto the primary.

This model has the advantage that it would naturally account for the
lack of optical polarisation. In addition, if the accretion stream
flow impacts with the primary over a range of azimuth, it is likely
that the brightness of the accretion site would not be uniform. This
would result in a non-symmetric X-ray light curve: this is consistent
with the X-ray light curves shown in Figure \ref{fold}. However, as
with the double degenerate polar model, emission lines (possibly weak)
would be expected to be observed in optical spectra.

After we submitted this paper, we became aware of a paper by Marsh \&
Steeghs (2002) who also propose this double-degenerate Algol model for
RX J1914+24.

\subsection{Neutron star-white dwarf pair}

We now consider a second alternative, that of a neutron star primary
and a white dwarf secondary. A number of accreting neutron stars are
known. These include low mass X-ray binary systems or isolated neutron
stars which accrete from the interstellar medium. The luminosity of an
accreting neutron star depends on the accretion rate, \Mdot. For an
isolated neutron star the accretion rate is determined by the proper
motion of the star, and the gas density in the interstellar medium.
(eg Treves et al 2000). Assuming a number density of interstellar gas
$\sim 0.1-1~{\rm cm}^{-3}$ and a gas temperature of $10^3-10^4$~K,
typical of the solar neighbourhood, $\dot M \sim 10^{11}-10^{12}~{\rm
g~s}^{-1}$, implying that the luminosity of an accreting isolated
neutron star would be \ltae $10^{32}~{\rm erg~s}^{-1}$.

Our estimate for the X-ray luminosity (\S \ref{extinc}) is a factor
$\sim$10 greater than this. However, if the neutron star is accreting
at a low rate, say from material located in the binary system
which maybe the remnants of a previous evolutionary phase, then the
predicted luminosity could equal the observed luminosity.

We conclude that a neutron star-white dwarf model for RX J1914+24 is
only viable if the neutron star is accreting material which has been
left over from a previous phase of binary evolution. This scenario is
attractive in the sense that we would not expect to observe
significant levels of polarisation in the optical band. Further, we
would not necessarily expect to observe optical line emission if the
neutron star was accreting spherically. In addition, if accretion was
occurring onto one of the magnetic poles of the neutron star the X-ray
light curve could match the observed X-ray profile for a wide range of
geometries. As in the double degenerate polar interpretation the hot
polar cap could irradiate the secondary white dwarf and cause a phase
shift in the X-ray/optical data.

\subsection{Unipolar-Inductor Model}   

In our double degenerate polar model the less massive degenerate star,
the secondary, must fill its Roche-lobe in order to facilitate the
mass-transfer process. Suppose that the companion is now slightly
denser and unable to fill its Roche lobe. The companion is still
tidally locked into synchronous rotation with the orbit, while the
magnetic primary may not be in perfect synchronism. The secondary will
then traverse the magnetic field of the primary. Because white
dwarfs are highly conductive, a large e.m.f. will be induced across
the secondary. If the space between the two white dwarfs is filled
with plasma, current loops can be set up (closed circuits), and the
system acts like a unipolar inductor. Dissipation occurs mainly in the
atmospheres of the white dwarfs, where the resistance is highest. We
now consider an alternative to our accretion driven scenarios.

One example of a cosmic unipolar inductor is a planet-moon system. A
well known example is the Jupiter-Io system, of which a bright trail
of foot-points of the magnetic-field lines traversed by Io are
observed on the surface of Jupiter (Clarke et al 1996). Another type
of unipolar inductor is a white dwarf-planet system. For such a system
with an orbital period of 10 hrs, electrical energy with a power $\sim
10^{29}$~erg~s$^{-1}$ (Li, Ferrario \& Wickramasinghe 1998) can be
generated. A double-degenerate unipolar inductor can produce a large
emf because of the strong magnetic field of primary dwarf and the
large diameter of the secondary white dwarf. Thus the electrical
current and hence the power that it generates should be significantly
larger than a planet-moon or a white dwarf-planet system. The details
of the unipolar-inductor model for double-degenerate star are
presented in Wu et al.\ (2002).

This model predicts X-ray emission from both magnetic poles. We would
expect the emission from each foot-point to result in light curves
which were `top-hat' in appearance. The X-ray light curves shown in
Figure 2 are consistent with a combination of two such light curves.
Further, if the foot-points of the current carrying field-lines remain
fixed with respect to the secondary (as in the Jupiter-Io system,
Clarke et al 1996) then asynchronous rotation will not manifest itself
in orbital period changes. The stability of the X-ray light curves is
consistent with this scenario.

Further, a luminosity of $10^{33}$ erg~s$^{-1}$ (cf \S \ref{extinc})
can be produced if the system deviates sufficiently from synchronism
(eg 1/1000; Wu et al, 2002).  As there is no accretion column and no
accretion shock, no strong emission lines or circular polarisation are
expected. Thus, the unipolar inductor can naturally account for the
two main difficulties with the accretion models.

\section{Conclusions}  

We have presented optical and infrared spectra of RX J1914+24: no
strong emission lines are present. Further, there is no evidence for
significant levels of polarisation. In the double degenerate polar
model, emission lines and detectable levels of polarisation are
expected. It is, however, possible that the emission lines are too
faint to have been detected in our spectra. We have also explored
other alternative models: a double degenerate Algol, a neutron
star-white dwarf pair and a cosmic unipolar inductor. The latter model
is attractive in that the non-detection of polarisation and emission
lines is a natural consequence of this model. It can also account for
the X-ray luminosity if the system has a small degree of asynchronism.

\section{Acknowledgments}

We would like to thank the NOT for the allocation of observing time.
NOT is operated on the island of La Palma jointly by Denmark, Finland,
Iceland, Norway, and Sweden, in the Spanish Observatorio del Roque de
los Muchachos of the Instituto de Astrofisica de Canarias.  We would
like to thank the staff of UKIRT for providing a spectrum which was
obtained as part of the UKIRT service program.  UKIRT is operated by
the Joint Astronomy Centre on behalf of the U.K. Particle Physics and
Astronomy Research Council. K.W. acknowledges the support of PPARC
for a Visiting Fellowship and the ARC Australian Research
Fellowship. Studies of magnetic stars and stellar systems at Steward
Observatory is provided by NSF grant AST97-30792 to G.D.S.

\end{document}